\documentclass[11pt,notitlepage]{article}
\usepackage{palatino} 
\usepackage[T1]{fontenc}
\usepackage{datetime}
\usepackage{lipsum} 
\usepackage{amsfonts, amsmath, amsthm, amssymb} 
\usepackage{graphicx} 
\usepackage{booktabs} 
\usepackage{wrapfig} 
\usepackage[labelfont=bf]{caption} 
\usepackage[top=1.0in,bottom=1.0in,left=1.0in,right=1.0in]{geometry} 
\usepackage{url}
\usepackage{abstract}
\usepackage{wrapfig}
\usepackage[T1]{fontenc}
\usepackage{array}
\usepackage{makecell}
\usepackage{tikz}
\usepackage{float}
\usepackage[utf8]{inputenc}
\usepackage{titlesec}
\newcommand*{\titleGP}{\begingroup 
\centering 
\vspace*{\baselineskip} 

\rule{\textwidth}{1.6pt}\vspace*{-\baselineskip}\vspace*{2pt} 
\rule{\textwidth}{0.4pt}\\[\baselineskip] 

{\Huge Exploring the Generality of a Java-based Loop Action Model for the Quorum Programming Language}

\rule{\textwidth}{0.4pt}\vspace*{-\baselineskip}\vspace{3.2pt} 
\rule{\textwidth}{1.6pt}\\[\baselineskip] 

\scshape 
\Large Prelim-Research Report

\vspace*{2\baselineskip} 

{\LARGE Preetha Chatterjee} 

{\itshape Department of Computer \& Information Sciences,

University of Delaware

\vspace{0.5cm}

\textup{Email: preethac@udel.edu}}

\vspace{0.5cm}

\vspace{0.5cm} 

\vspace{5cm}
\includegraphics[scale=0.5]{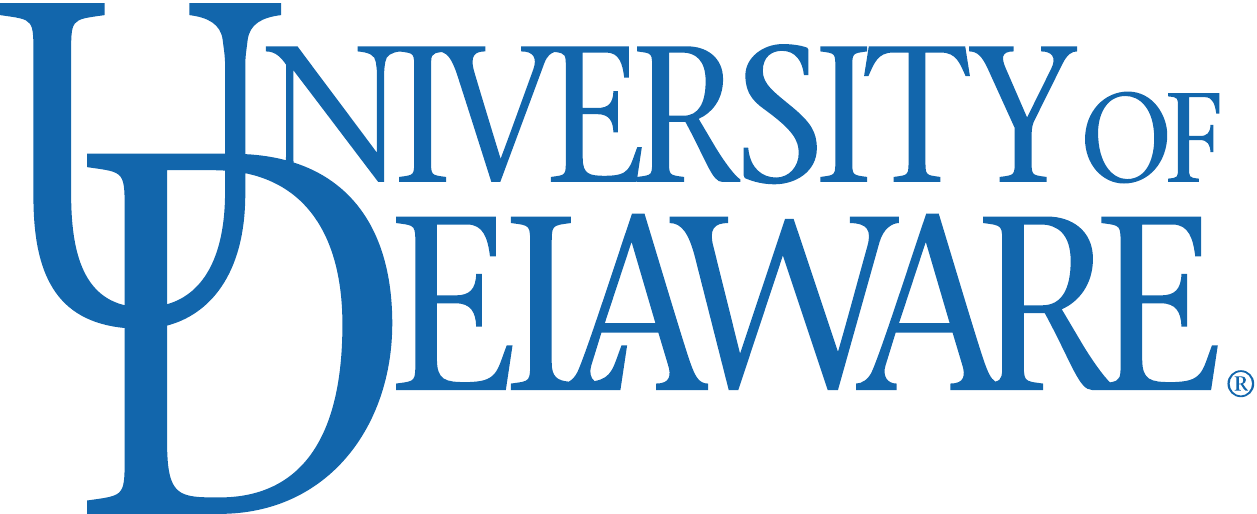}

\pagebreak
\endgroup}

\newcolumntype{x}[1]{>{\centering\arraybackslash}p{#1}}

\newcommand\diag[4]{%
  \multicolumn{1}{|p{#2}|}{\hskip-\tabcolsep
  $\vcenter{\begin{tikzpicture}[baseline=0,anchor=south west,inner sep=#1]
  \path[use as bounding box] (0,0) rectangle (#2+2\tabcolsep,\baselineskip);
  \node[minimum width={#2+2\tabcolsep},minimum height=\baselineskip+\extrarowheight] (box) {};
  \draw (box.north west) -- (box.south east);
  \node[anchor=south west] at (box.south west) {#3};
  \node[anchor=north east] at (box.north east) {#4};
 \end{tikzpicture}}$\hskip-\tabcolsep}}

\sffamily
\begin{document}

\titleGP
\tableofcontents
\newpage
\begin{center}
\LARGE{\textbf{Exploring the Generality of a Java-based Loop Action Model for the Quorum Programming Language}}
\end{center}
\renewcommand{\abstractnamefont}{\normalfont\large\bfseries}
\vspace{0.5cm}
\begin{abstract}
Many algorithmic steps require more than one statement to implement, but not big enough to be a method (e.g., add element, find the maximum, determine a value, etc.). These steps are generally implemented by loops. Internal comments for the loops often describe these intermediary steps, however, unfortunately a very small percentage of code is well documented to help new users/coders. As a result, information at levels of abstraction between the individual statement and the whole method is not leveraged by current source code analyses, as that information is not easily available beyond any internal comments describing the code blocks. 

Hence, this project explores the generality of an approach to automatically determine the high level actions of loop constructs.  The approach is to mine loop characteristics of a given loop structure over the repository of the Quorum language source code, map it to an (already developed for Java) action identification model \cite{XL15}, and thus identify the action performed by the specified loop. The results are promising enough to conclude that this approach could be applied to other programming languages too.
\end{abstract}

\section{Introduction}
In a software program, typically, multiple algorithmic steps combine to form a method. Examples are: Finding an element that satisfies some condition, comparing all pairs of corresponding elements from two collections. Although the algorithmic steps (which are the building blocks of the method) are small steps, they generally take more than one line of code to implement. The internal comments usually describe the actions of these algorithmic steps. However, descriptive internal comment, which would help the reader understand the code in a easier and better way, are very rare \cite{SN05}, \cite{KM01}. Adding to the lack of comments is the inadequacies of relying on names in the code. Not always can the method names and the variable names be defined in such a manner that the complete logic of the steps performed is subjectively clear from just ``reading the names''. 

\vspace{0.2cm}

Thus, information for the whole method, with details of individual algorithmic steps is not captured by present source code analyses. The main reason is the lack of internal comments. 
The current software tools performing source code analyses treat this problem in one of three ways. The first approach is to treat each method as a single unit, thus the source code analyses takes into account only the overall task performed by the method and not the detailed algorithmic steps which create it. The second approach is to treat the method as a set of individual statements. For the second approach, some documentation generators for method summaries process the methods as a set of individual statements and then select a subset of statements for which to generate a method summary \cite{sridhara2010towards}. The third approach is to use methods as a ``bag of words'' and select a subset of words for the summary \cite{5645482}. While Sridhara et.al \cite{sridhara2011automatically} used a set of templates to identify high level abstractions and generate summary comments for methods, Xiaoran et al. \cite{XL15} developed an action identification model to do the same without using manually created templates, and thus implemented a more flexible approach. 

\vspace{0.2cm}

This project focuses on exploring the generality of the Java-based Loop Action Model developed by Wang et.al \cite{XL15} for the Quorum Programming Language \cite{Quorum} . Quorum is a programming language designed specially for visually impaired middle and high school students. The objective is to investigate summarizing loops in Quorum, to identify the higher level abstraction of the action being performed by the loops.

\vspace{0.2cm}

The steps of this project are:
\begin{enumerate}
\item Manually identify loop-if structures (loops that contain exactly one conditional statement - ``if-statement'', which is also the last lexical statement within the loop body) from sample Quorum code, for at least 25 loops, and extract the corresponding feature vectors.
\item Manually map the identified loop-ifs with the already developed action identification model \cite{XL15} and validate the results.
\item Develop an automatic feature extraction system by using ANTLR(parser generator) \cite{TP13}.
\item Evaluate the output of the automatic tool/system.
\end{enumerate}

\vspace{0.2cm}

As Quorum is a comparatively new programming language, the repository of sample projects available for the purpose of research is small. For this project, the Quorum language compiler and its standard library files are used as the subjects of study. So understandably, the results could not be evaluated on different coding practices from different developers. To address that concern, after successful implementation of the project, an unbiased question-answer survey was conducted, to evaluate the correctness of the results.

\vspace{0.4cm}

\textbf{Contributions:}
The main contributions of the project include:
\begin{itemize}
\item an automatic tool to identify the high level actions implemented by Quorum loops.
\item demonstration of the feasibility of using the Java Loop Action Model to summarize loops in Quorum.
\item evaluation results from human judgement study that indicate strong positive opinion of the tool's effectiveness in automatically identifying high level actions for these loop structures.
\end{itemize}

Beyond showing generality of the loop action model, the perceived impact of the work is also to help blind programmers by providing them with the summary rather than reading the detailed loop code. This project has the potential to increase the effectiveness of code search tools and comment generator tools by providing the action phrase with the associated loop. It would also help to obtain a better comprehension of code, especially for blind readers.

\section{Background}
\subsection{Developing A Model of Loop Actions}
\begin{figure}[H]
 \begin{center}
    \includegraphics[width=83mm]{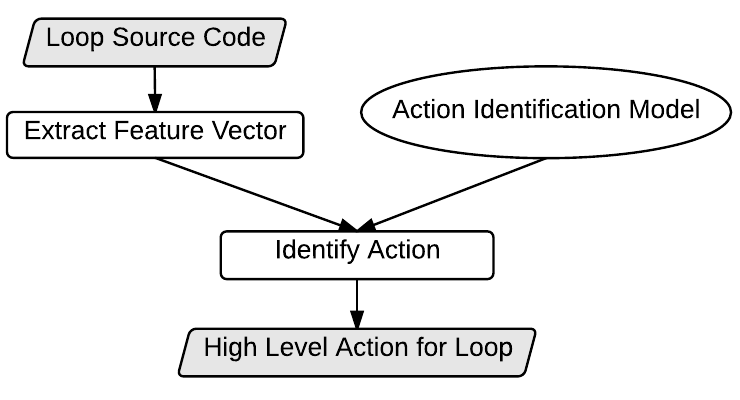}
  \end{center}
  \caption{Action Identification Process}
  \label{fig:action_iden_proc}
\end{figure}

Motivation for this project comes from existing work by Xiaoran Wang, Dr. Lori Pollock and Dr. Vijay Shanker on ``Developing a Model of Loop Actions by Mining Loop Characteristics from a Large Code Corpus'' \cite{XL15} - which involved identifying the higher level abstraction of the action being performed by a particular loop structure in Java based on their structure, data flow and linguistic characteristics. Their approach (Figure \ref{fig:action_iden_proc}) was to first identify action units (a code block that consists of a sequence of consecutive statements that logically implement a high level action) that are implemented by loop structures, characterize the loops as feature-value pairs to generate the loop feature vector (set or sequence of feature values) and then develop a model (action identification model) that can associate actions with loops based on their loop feature vectors.

\subsubsection{Terminology}
The following terminology is used to describe the features, that are used to determine the loop actions. This terminology was developed by Wang et al. \cite{XL15}.

\begin{figure}[H]
 \begin{center}
    \includegraphics[width=80mm]{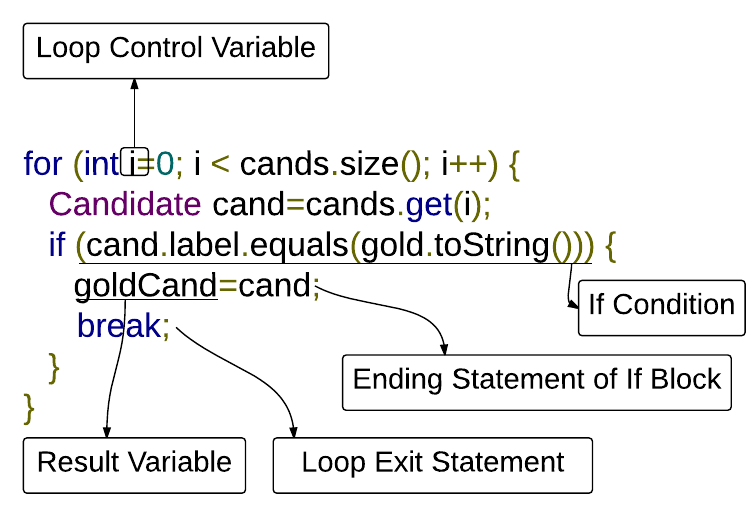}
  \end{center}
  \caption{Example of terminology}
  \label{fig:eg_term}
\end{figure}

Figure \ref{fig:eg_term} shows an example loop in Java to demonstrate the terminology used in developing the feature vector. 

\begin{itemize}
\item \textbf {If Condition:} The if condition refers to the conditional expression in the if statement of the loop.
\item \textbf{Loop Exit Statement:} Loop exit statements transfer control to another point in the code by exiting when control reaches the loop exit statement, such as a break or return.
\item \textbf {Ending Statement of if block:} As the last statement inside a loop-if is an if, the last executed statement of the loop is the last statement inside the if-block, which is referred as the ending statement. If the last executed statement of the loop is a branching statement like return, break or throw, then the statement immediately preceding the branching statement is considered to be the ending statement.
\item \textbf {Loop Control Variable:} The loop control variable is the variable defined in the loop condition.
\item \textbf {Result Variable:} Result variable captures the resulting value of the loop's action (if one exists). If the ending statement is an assignment, the result variable is the left-hand-side variable. If the ending statement is an object method invocation, it is the object that invokes the method.
\end{itemize}

\subsubsection{Features}
The features of loop-ifs separated into features related to ending statements and features related to the if condition are described \cite{XL15}.

\vspace{0.15cm}

 \underline {Features related to Ending Statement: }

\vspace{0.1cm}

\textbf{F1: Type of ending statement - }
If the last executed statement of the loop is a branching statement like return, break or throw, then the statement immediately preceding the branching statement is considered to be the ending statement.

\textbf{F2: Method name of ending statement method call - }
If the ending statement is a method invocation, the first verb comprising the method name is extracted. In table \ref{table:Semantic_feature}, they have however used method names occurring in the 100 most frequent loop-ifs.

\textbf{F3: Elements in collection get updated - }
F3=1 is set, when the result variable is the loop control variable; otherwise, F3=0.

\textbf{F4: Usage of loop control variable in ending statement - }
The loop condition determines the maximum number of iterations that will be executed. F4=0 is set, when the loop control variable never appears in the ending statement; F4=1, when the loop control variable is directly used in the ending statement; F4=2, when the loop control variable is on a def-use chain to a use in the ending statement.

\textbf{F5: Type of loop exit statement - }
F5 denotes if there is a control flow disruption in the action unit, and if there is, then the type (break, return, return boolean, return object or throw).

\vspace{0.3cm}

\underline {Features related to the if-condition:}

\vspace{0.1cm}

\textbf{F6: Multiple collections in if condition - }
This feature is a boolean that indicates whether multiple collections are compared in the if condition. F6=1 if there are two synchronized collections in the if condition; otherwise, F6=0.

\textbf{F7: Result variable used in if condition - }
F7=1 is set, if the result variable appears in the if condition; otherwise, F7=0.

\textbf{F8: Type of if condition - }
If the if condition is a numeric comparison( "<" , ">", "<=" or ">=" ) then F8=1 is set. Otherwise, if the type of if condition is a boolean value returned from a user defined method, then F8=2.

\vspace{0.5cm}

Table \ref{table:Semantic_feature} details the potential values for each loop-if feature.

\begin{table}[H]
\scriptsize
\begin{center}
\begin{tabular}{ |p{0.9cm}|p{5cm}|p{9cm}| } 
 \hline
 \textbf {Label} & \textbf{Feature} & \textbf{Possible Values and their Semantics} \\ 
 \hline
 F1 & Type of ending statement & 0: none 1: assignment 2: increment 3: decrement 4: method invocation 5: object method invocation 6: boolean assignment \\ 
 \hline
 F2 & Method name of ending statement method call & 0: none 1: add 2: addX 3: put 4: setX 5: remove \\ 
 \hline
 F3 & Elements in collection get updated & 0: false 1: true\\
 \hline
 F4 & Usage of loop control variable in ending statement & 0: not used 1: directly used 2: used indirectly through data flow\\
 \hline
 F5 & Type of loop exit statement & 0: none 1: break 2: return 3: return boolean 4: return object 5: throw\\
 \hline
 F6 & Multiple collections in if condition & 0: false 1: true\\
 \hline
 F7 & Result variable used in if condition & 0: false 1: true\\
 \hline
 F8 & Type of if condition & 1: >/</>=/<=   2: others\\
 \hline
\end{tabular}
\caption{Semantics of Feature Values}
\label{table:Semantic_feature}
\end{center}
\end{table}

\textbf{From Loop to Feature Vector: An example}

\vspace{0.2cm}

The feature vector for the example code fragment in Figure 2 is:(F1:1, F2:0, F3:0, F4:2, F5:1, F6:0, F7:0, F8:2). 
F1 indicates that the ending statement is an assignment. 
F2 indicates there is no method name from an ending method call. 
F3 indicates that not every element in the collection is updated. 
F4 indicates that the loop control variable is on the def-use chain to a use in the ending statement. 
F5 indicates that the type of loop exit statement is a break. 
F6 indicates that there is only one collection in if condition. 
F7 indicates that the result variable is not used in the if condition. 
F8 indicates that the type of the if condition is not numeric comparison.

\vspace{0.4cm}

\subsubsection{Action Identification Model}

 \setlength\extrarowheight{4pt}
\begin{table}[H]
\scriptsize
\begin{center}
\begin{tabular}{|c|c|c|c|c|c|c|c|c|}\hline
\diag{0.175em}{2.5cm}{ Action }{ Feature }& F1 & F2 & F3 & F4 & F5 & F6 & F7 & F8\\
 \hline
 count & 2 & & & & 0 & & & \\ 
 \hline
  determine & 0 & & &  & 2,3 & & & \\ 
 \hline
  determine & 6 & & &  & 0,1 & & & \\ 
 \hline
  max/min & 1 & & & 1,2 & 0 & 0 & 1 & 1 \\ 
 \hline
  find & 1 & & & 1,2 & 1,2,4 & & & \\ 
 \hline
  find & 0 & & & & 4 & & & \\ 
 \hline
  copy & 5 & 1 & 0 & 1 & 0 & & & \\ 
 \hline
  ensure &  & & & & 5 & & & \\ 
 \hline
  compare &  & & & & 3 & 1 & & \\ 
 \hline
  remove & 5 & 5 & 1 & 1 & 0 & & & \\ 
 \hline
  get & 5 & 1,3 & 0 & 2 & 0 & & & \\ 
 \hline
  add & 5 & 2 & 0 & 1,2 & 0 & & & \\ 
 \hline
  set\_one  & 5 & 4 & 0 & 1,2 & & & & \\ 
 \hline
  set\_all & 5 & 4 & 1 & 1,2 & 0 & & & \\ 
 \hline
\end{tabular}
\caption{Action Identification Model}
\label{table:action_id_model}
\end{center}
\end{table}

\begin{figure}[H]
\begin{center}
\includegraphics[width=160mm]{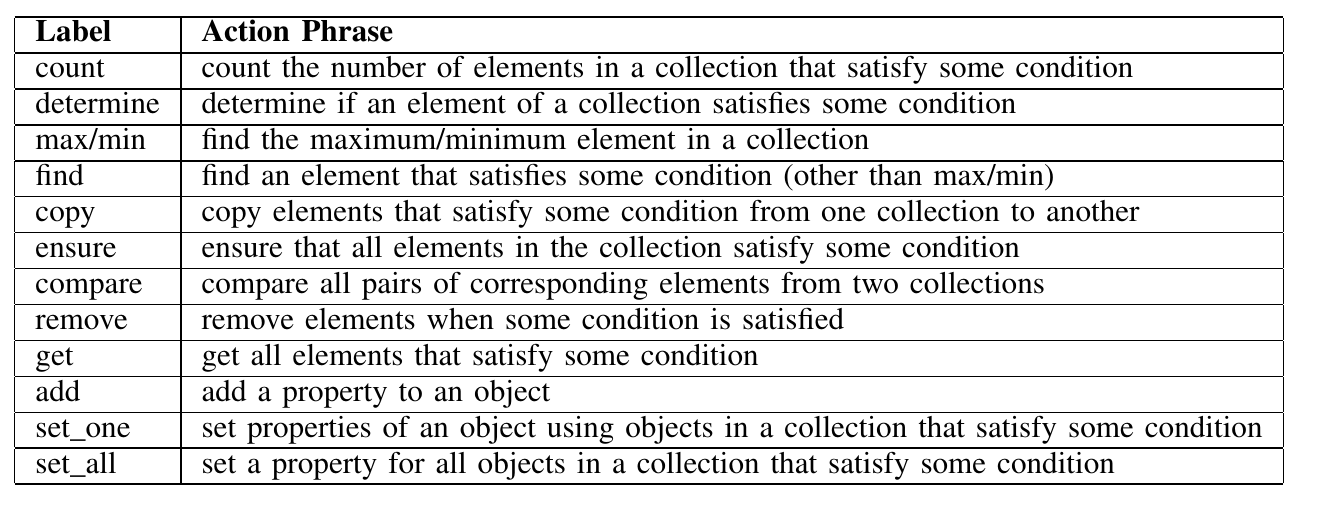}
\end{center}
\caption{Actions with their verb phrase descriptions}
\label{fig:phrase}
\end{figure}

To characterize the high level action performed by a specific feature vector, Wang et al. \cite{XL15} examined several loops corresponding to that loop feature vector that had comments associated with them, extracted the verbs from comments, computed verb distributions for vectors, clustered vectors based on verb distribution and selected representative action for each cluster. Thus, they developed an action identification model, where each row shows an action and its corresponding set of feature values, as shown in Table \ref{table:action_id_model}. For example, if a loop has value 0 for F1 and 2 or 3 for F5, or value 6 for F1 and 0 or 1 for F5, then the model will label this loop with the action ``determine''. Figure \ref{fig:phrase} shows the action phrases for each of the actions.

\vspace{0.2cm}

\subsubsection{Limitations}

The authors note several limitations of the work so far: 
\begin{itemize}
\item The only loop formats considered in Java were: for, enhanced-for, while or do-while.
\item The paper focused on loops that contain exactly one conditional statement (if-statement), which is also the last lexical statement within the loop body (a loop-if structure). 
\item Nested loops were not in the scope. 
\item The action identification table was developed based on the 100 most frequent loop feature vectors in the data set used for the project, and not for all possible loop feature vectors.
\end{itemize}

\subsection{Quorum Language}
\subsubsection{Introduction}
Quorum is a programming language that is built, keeping in the mind the problems faced by the blind students to learn and use computer programming languages in general. To quote Dr. Andreas Stefik, one of the inventors of Quorum, "The blind and visually impaired community is significantly underrepresented in computer science. Students who wish to enter the discipline must overcome significant technological and educational barriers to succeed.  While much work has been dedicated to helping the blind use various computer technologies, more research is needed on finding ways to make it easier for blind users to obtain high-paying and meaningful careers. Indeed, with 61\% of working adults (aged 16 to 64) with vision loss out of the work force, and with households that include a blind member having a significantly higher rate of poverty, creating more opportunities for this group of individuals is sorely needed. "\cite{SH11}. 

\vspace{0.2cm}

Quorum started as an interpreted language originally designed to be easier to hear through screen readers for blind or visually impaired users. Eventually, it became a general purpose programming language designed for any user. Current versions compile to Java Bytecode and run on the Java Virtual Machine, similar to JRuby, Jython, or Scala. Quorum 3.0 also compiles to JavaScript and can be run from the web \cite{Quorum}. 

\subsubsection{Language Basics}
Quorum is an object-oriented programming language which has a general purpose type system, with generics for containers (e.g., arrays, hash tables, lists). Quorum also has a standard library, which contains many additions to the language, including math libraries, web components, and a game engine \cite{Quorum}. 

\vspace{0.2cm}

%
%

As this project explores whether the Java loop action model can be used for Quorum, it is important to understand the loop structures in Quorum, which are different from that of Java. First there is no for-loop. Instead there are three different loop types as follows:
\begin{itemize}
\item repeat <expression> times:
\small\begin{verbatim}
   integer a = complicatedMathAction()
   integer b = anotherComplexAction()
   repeat (b / a - (b + 5)) times
   end
\end{verbatim}
\normalsize

\item repeat while <expression> 
\small\begin{verbatim}
   integer a = 0
   repeat while a < 15
      a = a + 1
   end
 \end{verbatim}
\normalsize

\item repeat until <expression> 
\small\begin{verbatim}
    integer a = 0
    repeat until a < 15
       a = a + 1
    end
\end{verbatim}
\normalsize

In this case, since a is less than 15 this loop will execute 0 times. 

\end{itemize}

%
%
%
%
%
%

Source Code for the Quorum project can be found at the Quorum Bitbucket page \cite{Quorum_Repo} at:

 \url{https://bitbucket.org/stefika/quorum-language}. 

\section{Building a Loop Action Model for Quorum}
\subsection{Overview}
\begin{figure}[H]
 \begin{center}
    \includegraphics[width=140mm]{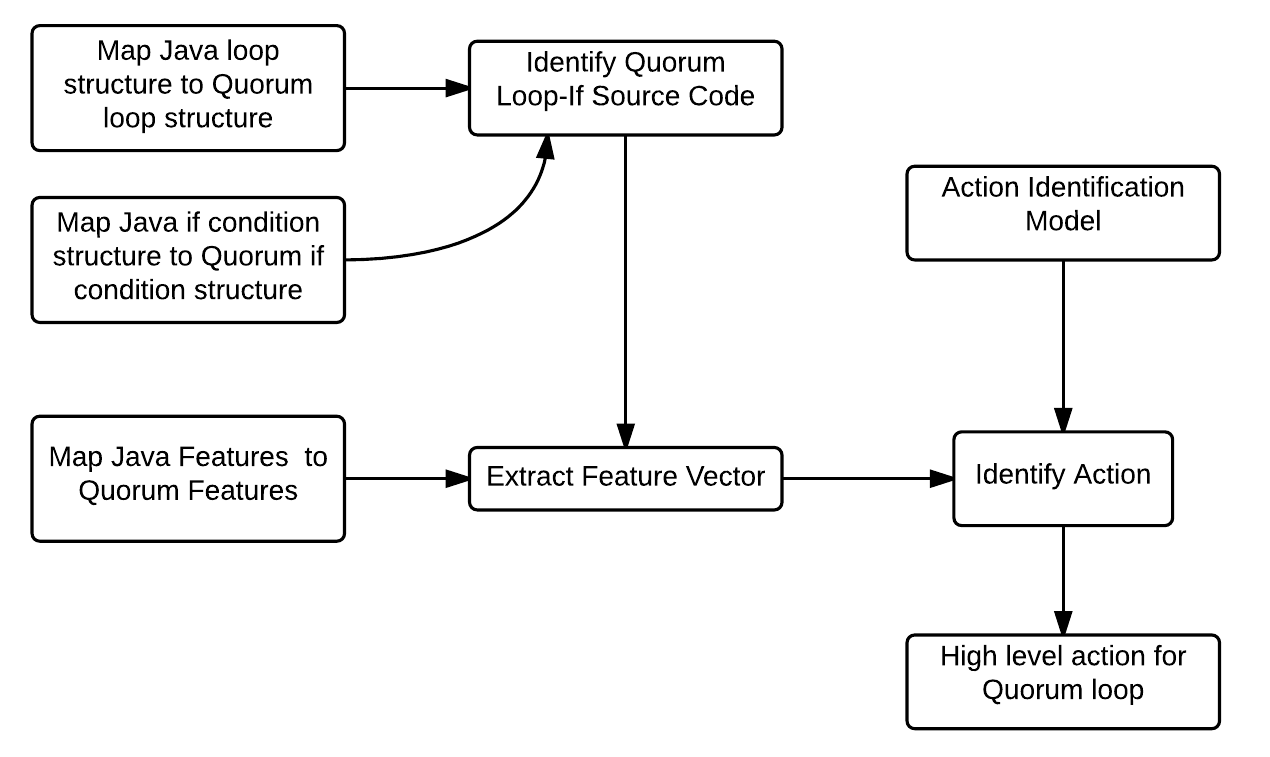}
  \end{center}
  \caption{From Java to Quorum}
  \label{fig:Action_id_proc}
\end{figure}

The goal of this project is to explore the generality of the Loop Action Model \& Feature Vector approach to identify high level actions for loop-ifs in Quorum. For Java, the action identification model is in table \ref{table:action_id_model}. The overall process of this task is shown in Figure \ref{fig:Action_id_proc}. To investigate generality to Quorum, we need to investigate whether and how the same loop features can be extracted from the Quorum loop-ifs and whether the same identification model is applicable in Quorum codes.

\vspace{0.2cm}

Thus, we need to map loop formats in Java (for, enhanced-for, while or do-while) to loop formats in Quorum (repeat times, repeat while and repeat until), if conditions in Java to if-conditions in Quorum and each feature value in Java to the equivalent in Quorum. After the mappings, Quorum Loop-if source code can be identified. The action identification model is then referenced to determine the high level action associated with the loop's feature vector.

\vspace{0.2cm}

\subsection{Manual identification of loop-if and mapping to Action Identification Model}

\subsubsection{Mapping Java loop structure to Quorum loop structure}
Quorum has no ``for'' or ``enhance-for'' loops. Instead the loops are of types: ``repeat <expression> times'', ``repeat while <expression>'' and ``repeat until <expression>'', all of which are considered in this project. 
\vspace{0.2cm}

\begin{itemize}
\item {\bf{repeat <expression> times:}}

There is no repeat <expression> times loop in Java. The code snippets below show the same logic written in Java (using a for-loop) and Quorum. Differences are bolded in each of them. The variable declaration and print statements are almost similar in both of the languages, but the loop structures are entirely different in this case.

\begin{figure}[H]
 \begin{center}
    \includegraphics[width=62mm]{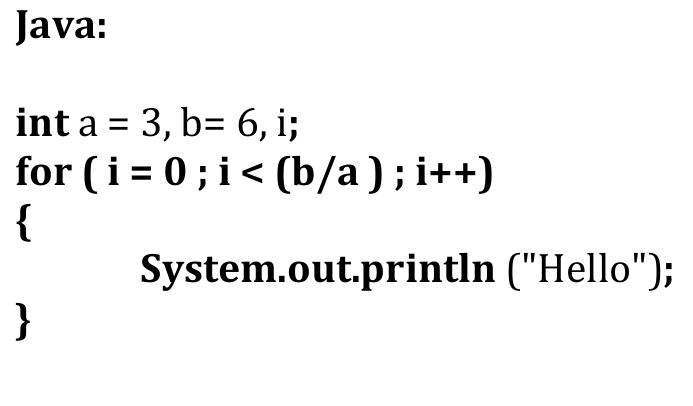}\hspace{2.5cm}
    \includegraphics[width=40mm]{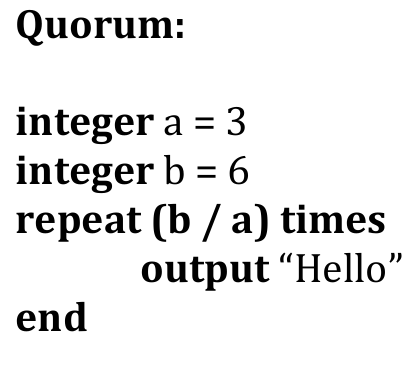}
  \end{center}
  \caption{Comparison of repeat <expression> times loop structures in Java (Left), and Quorum (Right)}
  \label{fig:rep_times}
\end{figure}

\item {\bf{repeat while <expression>:}}
Repeat while <expression> loops in Quorum are quite similar to while loops in Java. The code snippets below show the same logic written in Java (using a while-loop) and Quorum. The only major difference is the loop-syntax (``()'', ``{}", keywords ``repeat'' and ``end"), bolded below.

\begin{figure}[H]
 \begin{center}
    \includegraphics[width=52mm]{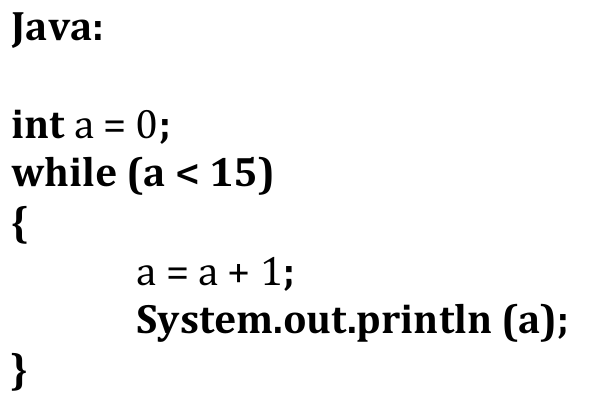}\hspace{2.5cm}
    \includegraphics[width=38mm]{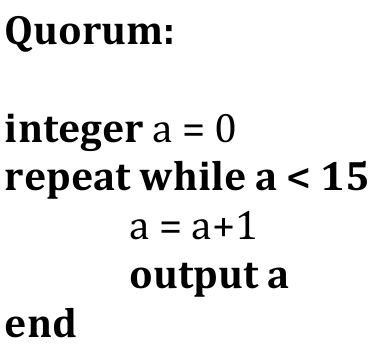}
  \end{center}
  \caption{Comparison of repeat while <expression> loop structures in Java (Left), and Quorum (Right)}
  \label{fig:rep_while}
\end{figure}

\item {\bf{repeat until <expression>:}}
There is no repeat until <expression> loop in Java. The code snippets below show the same logic written in Java (using a for-loop) and Quorum. Differences are bolded in each of them. As before, the loop structures are entirely different in this case as well.

\begin{figure}[H]
 \begin{center}
    \includegraphics[width=49mm]{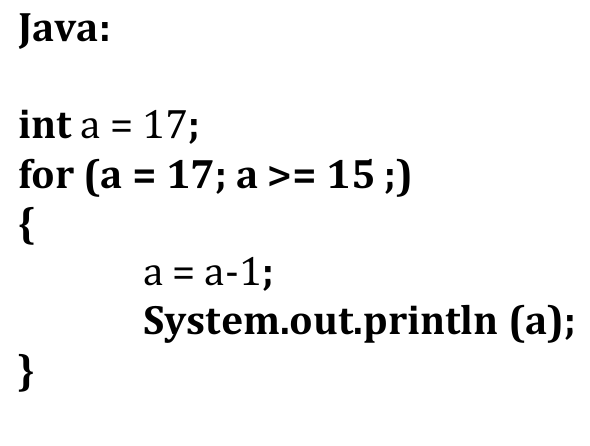}\hspace{2.5cm}
    \includegraphics[width=34mm]{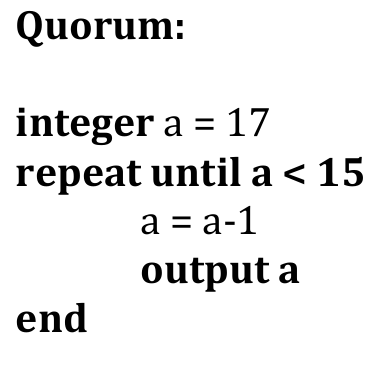}
  \end{center}
  \caption{Comparison of repeat until <expression> loop structures in Java (Left), and Quorum (Right)}
  \label{fig:rep_until}
\end{figure}

\end{itemize}

\subsubsection{Mapping Java if condition to Quorum if-condition structure}
The ``if conditional'' in Quorum is almost similar syntactically with ``if condition'' in Java. So mapping Java if-conditions to Quorum if-conditionals is straight forward. The code snippets below show the same logic written in Java and Quorum. The only differences are the use of ``()'' , ``;'' and ``end".

\begin{figure}[H]
 \begin{center}
    \includegraphics[width=30mm]{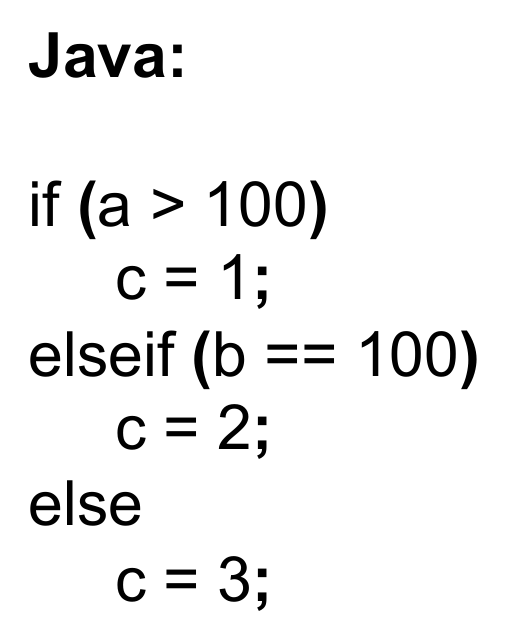}\hspace{2.4cm}
    \includegraphics[width=24mm]{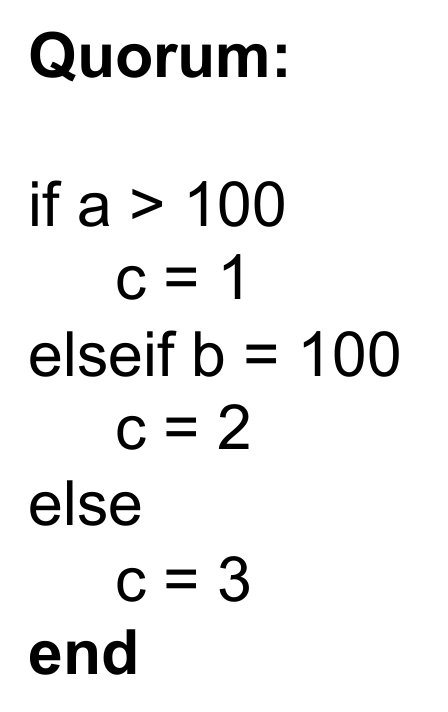}
  \end{center}
  \caption{Comparison of if structures in Java (Left), and Quorum (Right)}
  \label{fig:eg_if}
\end{figure}

\subsubsection{Mapping Java loop-if structure to Quorum loop-if structure}

\begin{figure}[H]
 \begin{center}
    \includegraphics[width=68mm]{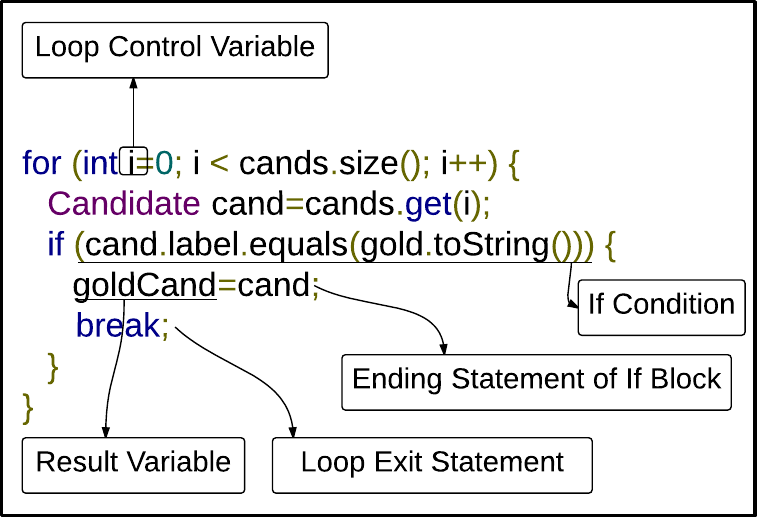}\hspace{1cm}
    \includegraphics[width=85mm]{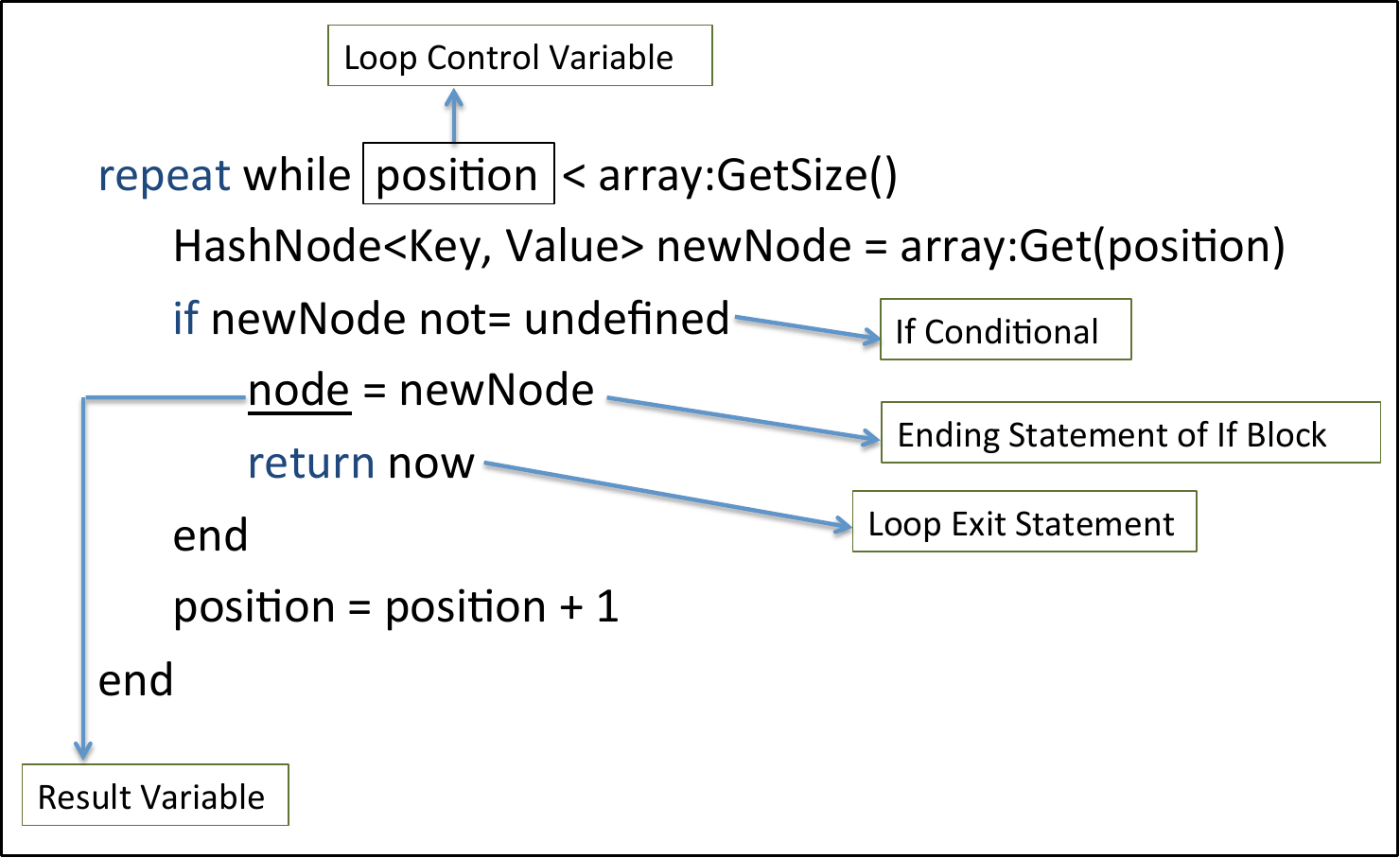}
  \end{center}
  \caption{Comparison of Loop-if structures in Java (Left), and Quorum (Right)}
  \label{fig:eg_term1}
\end{figure}

Figure \ref{fig:eg_term1} compares sample loop-if structures in Java and Quorum. The types of loop structures used in the examples in figure \ref{fig:eg_term1}, are different - Java (for-loop) and Quorum (repeat while <expression>). Hence, the Loop Control Variable (`i') in the for-loop in Java is mapped to the Loop Control Variable (`position') in the <expression> for Quorum. ``If-conditions", ``Ending Statement of if-block", ``Loop Exit Statement" and "Result Variable" are similar in both Java and Quorum .The only syntactic differences are in the use of ``()", ``\{\}", ``;", ``:", ``." and keyword ``end". 
\vspace{0.2cm}

However, within the Quorum repeat loop, there is an extra lexical statement right after the if-conditional: ``position = position + 1''. As per the definition, a loop-if structures in Java is a loop that contain exactly one conditional statement (if-statement) , which is also the last lexical statement within the loop body. But if a closer look is taken into the Quorum code snippet, it is clear that ``position" being the loop-control variable, the statement after the if-conditional is doing nothing else, but actually increasing the counter of the repeat loop, same as what ``i++" is doing in the Java for-loop. So in this example, though the Quorum and Java code snippets are syntactically different, but are semantically both ``loop-if" structures. 

\subsubsection{Mapping Java Feature Vectors to Quorum Feature Vectors}
The features of Java loop-ifs, described in section 2.1.2, are now mapped to that of Quorum.

\begin{itemize}

\item \textbf{F1: Type of ending statement - }
Similar to Java, the different types of ending statements (if present) are: assignment, increment, decrement, method invocation, object method invocation and boolean assignment. Thus this feature is identified the same in Java and Quorum.The syntactic type of ending statement can be a strong indicator of the overall purpose of the loop. If there is no ending statement in the loop, F1 is set to 0: none.

\item \textbf{F2: Method name of ending statement method call - }
Similar to Java, if the ending statement is a method invocation, the first verb comprising the method name is extracted. Thus this feature is identified the same in Java and Quorum. The values for this feature are - none, add, addX, put, set and remove.

\item \textbf{F3: Elements in collection get updated - }
Similar to Java, F3=1 is set, when the result variable is the loop control variable; otherwise, F3=0. Thus this feature is identified the same in Java and Quorum.

\item \textbf{F4: Usage of loop control variable in ending statement - }
Similar to Java, F4=0 is set, when the loop control variable never appears in the ending statement; F4=1, when the loop control variable is directly used in the ending statement; F4=2, when the loop control variable is on a def-use chain to a use in the ending statement. Thus this feature is identified the same in Java and Quorum.

\item \textbf{F5: Type of loop exit statement - }
F5 denotes if there is a control flow disruption in the action unit, and if there is, then the type. For Quorum, we have the following types - return, return boolean, return object. \textbf{Type4 or ``break" and Type5 or ``throw" in Java do not exist in Quorum}, hence they are not  possible values here.

\item \textbf{F6: Multiple collections in if condition - }
\textbf{Collections in Java} is equivalent to \textbf{Containers in Quorum}. This feature is a boolean that indicates whether multiple collections (containers) are compared in the if condition. F6=1 if there are two synchronized containers in the if condition; otherwise, F6=0.

\item \textbf{F7: Result variable used in if condition - }
Similar to Java, F7=1 is set, if the result variable appears in the if condition; otherwise, F7=0. Thus this feature is identified the same in Java and Quorum.

\item \textbf{F8: Type of if condition - }
Similar to Java, if the if condition is a numeric comparison( "<" , ">", "<=" or ">=" ) then F8=1 is set. Otherwise, for eg. if the type of if condition is a boolean value returned from a user defined method, then F8=2. Thus this feature is identified the same in Java and Quorum.

\end{itemize}

To summarize, with the slight modifications as mentioned above, only features F5 and F6 change for Quorum.  The modified table from Wang et al. \cite{XL15} for "Semantics of Feature Values" for Quorum is as follows:

\begin{table}[H]
\scriptsize
\begin{center}
\begin{tabular}{ |p{0.9cm}|p{5cm}|p{9cm}| } 
 \hline
 \textbf {Label} & \textbf{Feature} & \textbf{Possible Values and their Semantics} \\ 
 \hline
F1 & Type of ending statement & 0: none 1: assignment 2:increment 3:decrement 4:method invocation 5:object method invocation 6: boolean assignment \\ 
 \hline
F2 & Method name of ending statement method call & 0:none 1:add 2:addX 3:put 4:setX 5:remove \\ 
 \hline
F3 & Elements in collection get updated & 0: false 1: true\\
 \hline
F4 & Usage of loop control variable in ending statement & 0: not used 1:directly used 2:used indirectly through data flow\\
 \hline
 \textbf{F5} & \textbf{Type of loop exit statement} & \textbf{0:none 2:return 3:return boolean 4:return object} \\
 \hline
 \textbf{F6} & \textbf{Multiple collections(containers) in if condition} & \textbf{0: false 1: true}\\
 \hline
F7 & Result variable used in if condition & 0: false 1: true\\
 \hline
F8 & Type of if condition & 1: >/</>=/<=   2: others\\
 \hline
\end{tabular}
\caption{Semantics of Feature Values for Quorum (Differences from Java are bolded)}
\label{table:Semantic_feature_Quorum}
\end{center}
\end{table}

\subsubsection{From Loop to Feature Vector: An example}
A feature vector for a given loop-if is constructed by extracting the features F1 through F8 from the loop`s source code representation using simple static analysis. The feature vector for the example code fragment of Quorum in Figure \ref{fig:eg_term} is:
(F1:1, F2:0, F3:0, F4:2, F5:4, F6:0, F7:0, F8:2). 

\vspace{0.2cm}

F1 indicates that the ending statement is an assignment. F2 indicates there is no method name for an ending statement method call. F3 indicates that no element in a collection is updated. F4 indicates that the loop control variable is on indirect use in the ending statement. F5 indicates that the loop exit statement is returning an object. F6 indicates that there is no collection in if condition. F7 indicates that the result variable is not used in the if condition. F8 indicates that the type of the if condition is not numeric comparison. 

\vspace{0.2cm}

Mapping the feature vector to the action identification model in Table \ref{table:action_id_model}, this loop action is identified as `find'. Thus, the action can be identified as "find an element that satisfies some condition".

\subsubsection{Evaluation}
After the initial phase of manually identifying loop-if structures in Quorum and mapping to the Action Identification Model in Table \ref{table:action_id_model}, a study was conducted to evaluate the accuracy of this approach for Quorum. A set of 10 sample Quorum code snippets containing loop-if structures, was given to Xiaoran Wang, the first author of the paper - "Developing a Model of Loop Actions by Mining Loop Characteristics from a Large Code Corpus" \cite{XL15}. Loops were randomly selected - 60\% of sample snippets were loop-ifs for which actions could be identified by the approach. The evaluator was asked to determine the corresponding feature vectors and the identified actions. His expert results were compared with the results from manually applying the approach, and there was a 100\% match of the results.

\subsection{Developing a tool for Automatic Action Identification}

\begin{figure}[H]
 \begin{center}
    \includegraphics[width=185mm]{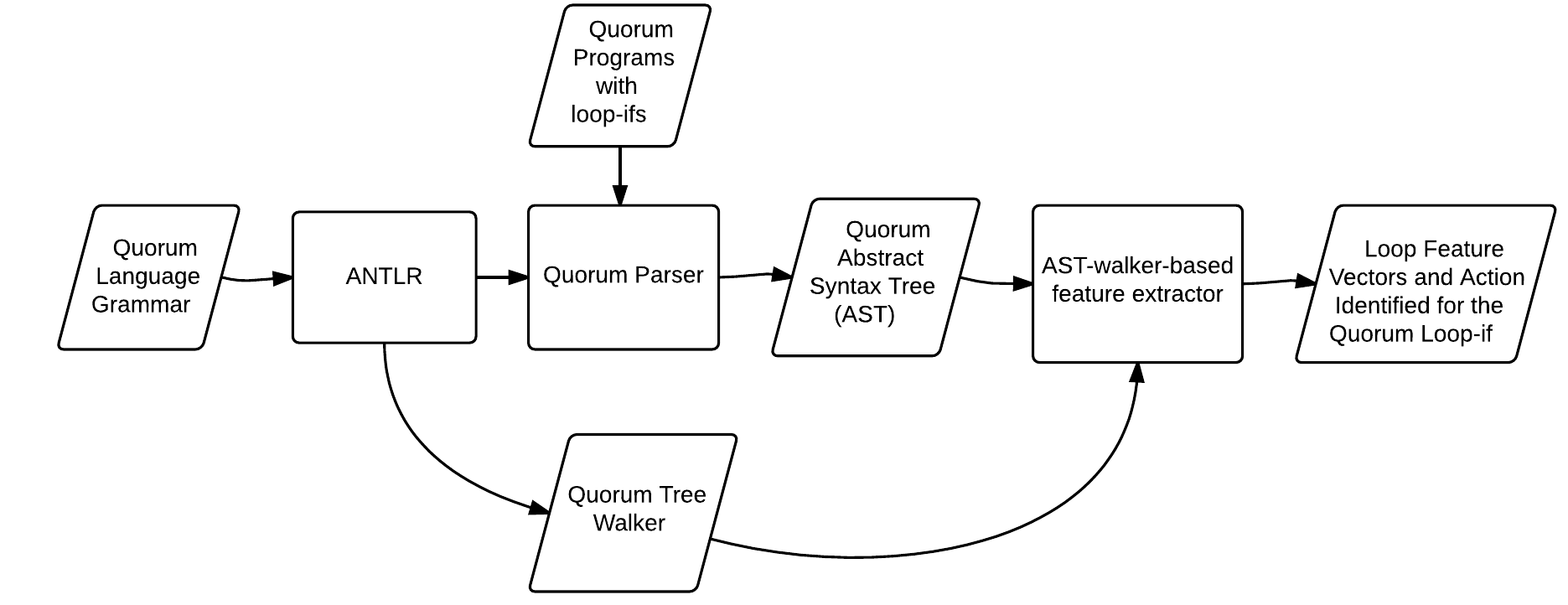}
  \end{center}
  \caption{Automatic Action Identification of a Quorum loop-if}
  \label{fig:Quorum_auto_action_id}
\end{figure}

The objective of developing a tool for Automatic Action Identification is to determine if given a piece of Quorum source code (containing a loop-if) as input, the tool is able to extract the corresponding feature vector values and identify the high level action performed by the loop. The overall approach to automatic action identification is shown in figure \ref{fig:Quorum_auto_action_id}.

%


\subsubsection{Quorum Language Grammar}
\begin{figure}[H]
 \begin{center}
    \includegraphics[width=65mm]{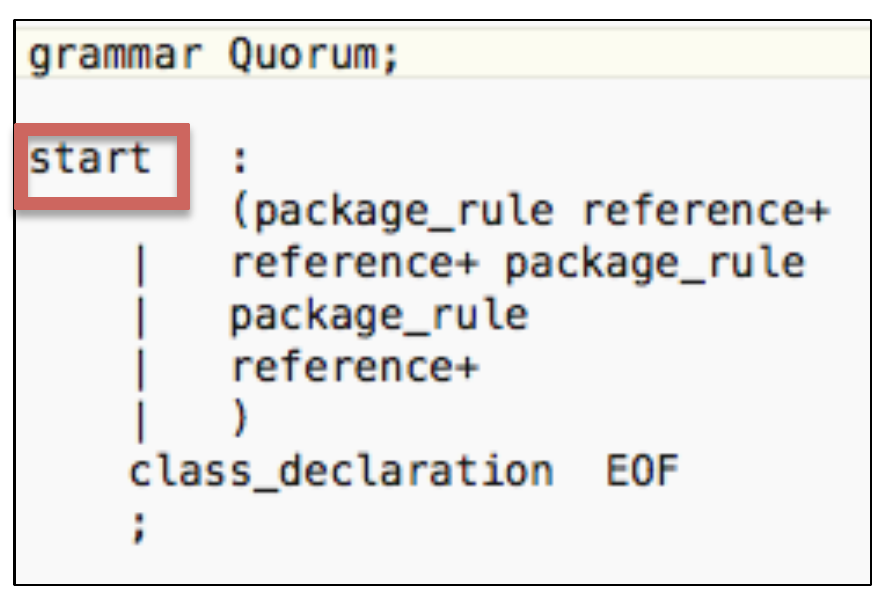}
     \includegraphics[width=65mm]{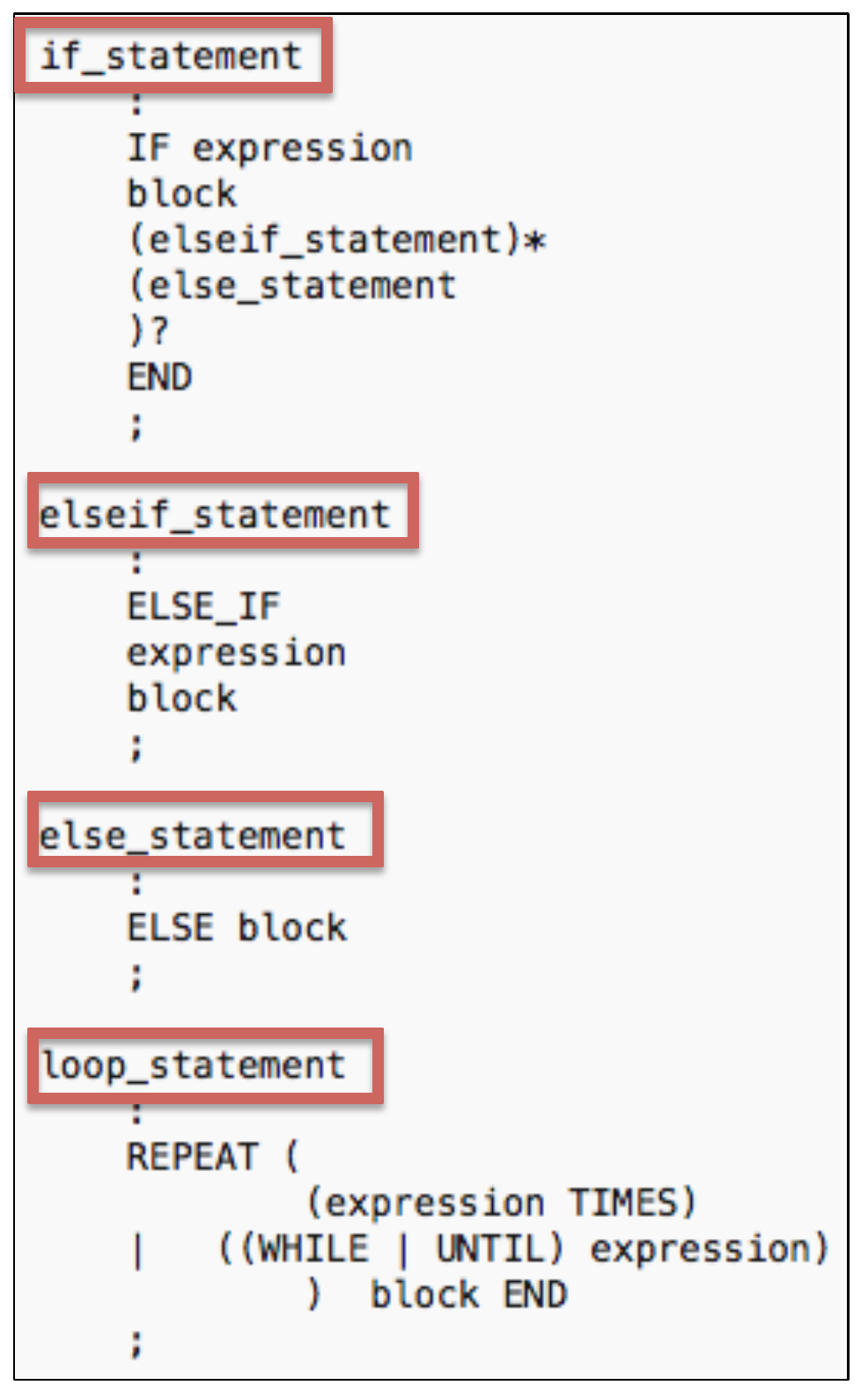}
      \includegraphics[width=155mm]{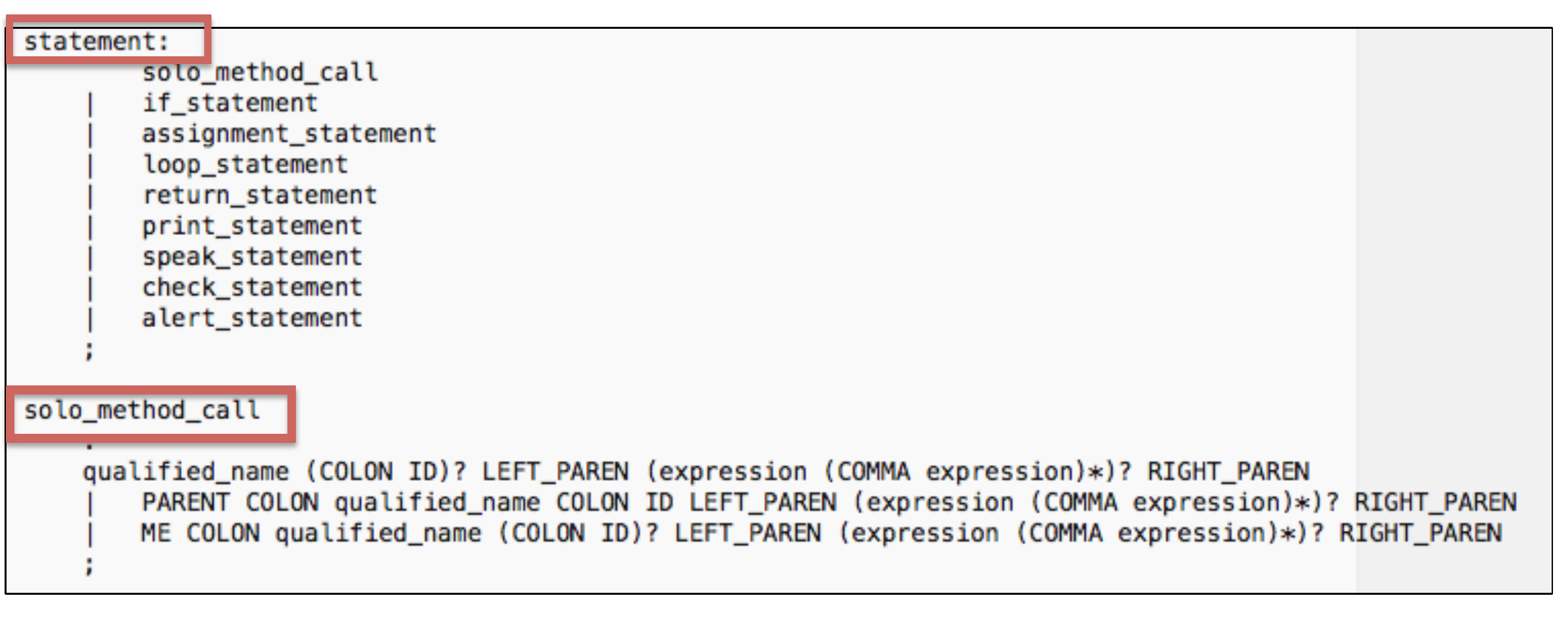}
  \end{center}
  \caption{Relevant Parts of Quorum Grammar}
  \label{fig:Quorum_grammar}
\end{figure}

A grammar formally defines the syntax rules of a language. The first step in the process for automation was to understand the basic architecture of the Quorum compiler and identify/extract Quorum language grammar from the source code repository of Quorum at \cite{Quorum_Repo}. Figure \ref{fig:Quorum_grammar} shows some of the relevant parts(rules) of the Quorum grammar referred to extract the feature vectors of loop-ifs.

\subsubsection{ANTLR}
ANTLR (ANother Tool for Language Recognition) is a parser generator for reading, processing, executing, or translating structured text or binary files. The latest version of the Quorum compiler uses ANTLR4 backend. From the Quorum language grammar, ANTLR generates a Quorum Parser that automatically builds Quorum Abstract Syntax Trees (AST) \cite{TP13}. ANTLR also automatically generates Quorum tree walkers that are used to visit the nodes of the ASTs. 

\subsubsection{Quorum Abstract Syntax Tree (AST)}


\begin{figure}[H]
 \begin{center}
    \includegraphics[width=185mm]{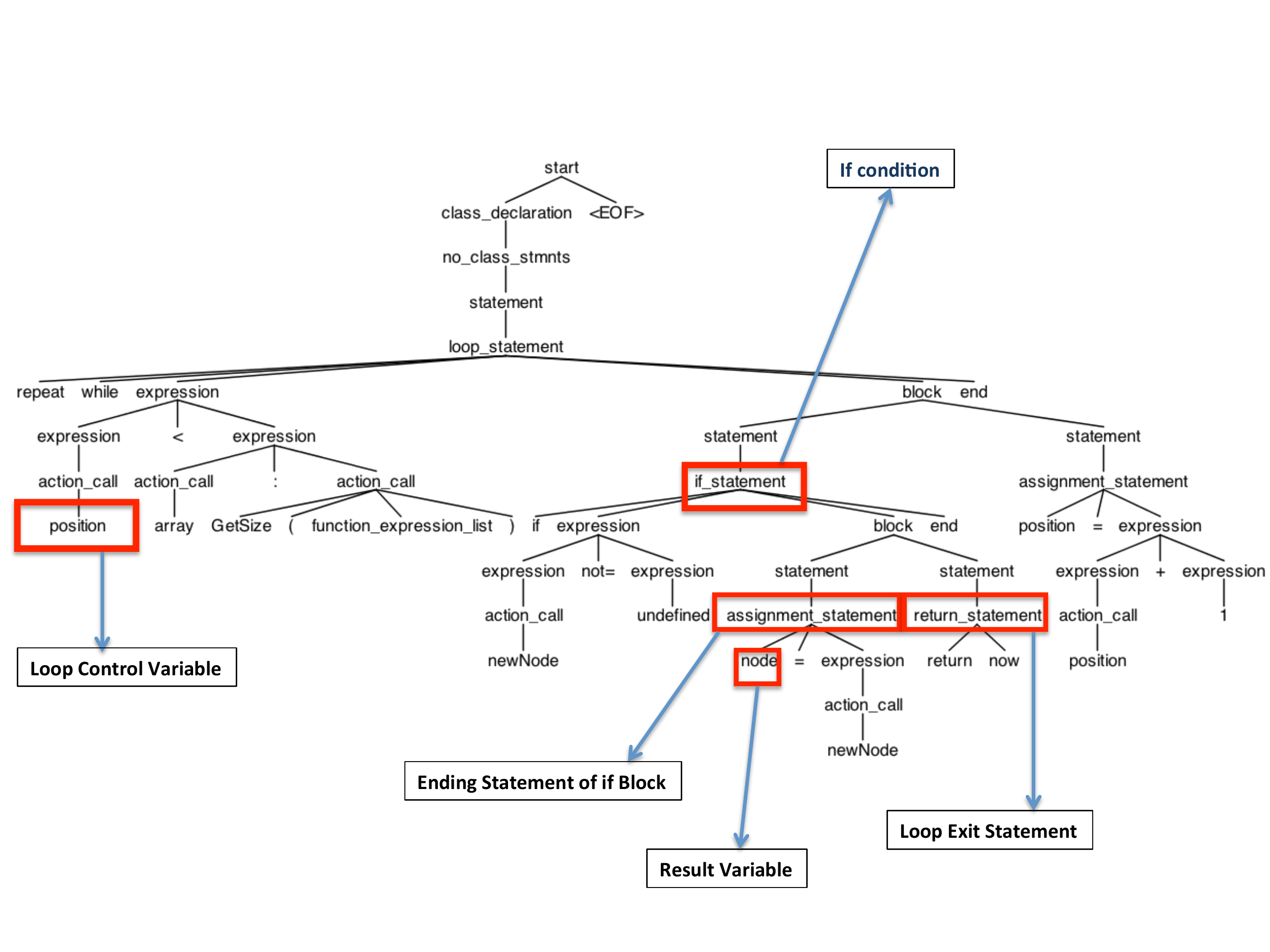}
  \end{center}
  \caption{Quorum Parse Tree for an example Quorum loop-if}
  \label{fig:Quorum_parse_tree_gui}
\end{figure}

Figure \ref{fig:Quorum_parse_tree_gui} shows the ANTLR generated parse tree for the Quorum (loop-if) code snippet in Figure \ref{fig:eg_term1}. The terminologies used for identifying the feature vectors of a loop-if (section 2.1.1) are highlighted in the figure. Each of the terminologies is identified in the context of the ASTs as follows: 
 
\begin{itemize}

\item \textbf{Loop Control Variable:} The Quorum code \textbf{repeat while position < array:GetSize} uses a repeat while <expression> loop. So, the loop-control variable ``Position" is to be located under the third child node ``expression" under parent node ``loop\_statement". The first and second child nodes of ``loop\_statement" are ``repeat" and ``while" respectively.

\item \textbf{If Condition:} We are only interested in if conditions inside the loop structure (i.e. a loop-if structure). So, an ``if-condition" is to be located under the fourth child node ``block" under parent node ``loop\_statement". Then, the second last node (as the last node pertains to the increment of loop control variable in this case) is examined for it's type . If it is an ``if\_statement", the particular Quorum code snippet under consideration \textbf{is identified to be a ``loop-if " structure}.

\item \textbf{Loop Exit Statement:} The Loop Exit Statement (if exists) should be the last statement inside the ``if block". So the parse tree is traversed down the path from the ``if\_statement" node to its child node ``block", down to the last ``statement node" under it. This ``statement" node is further investigated to identify it's type. If it is a ``return\_statement", it is concluded that the loop-if structure under consideration contains a ``Loop Exit Statement".

\item \textbf{Ending Statement of if Block:} The Ending Statement of if block is essentially the last statement inside the ``if block". However if a loop exit statement is present, the ending statement would be the second last statement inside the block (same as shown in Figure \ref{fig:Quorum_parse_tree_gui}). So the parse tree is traversed down the path from the ``if\_statement" node to its child node ``block", down to the second last ``statement" node under it. The type of an Ending Statement could be ``assignnment\_statement" (as shown above) or it could be a ``method\_call", depending on the code snippet examined.

\item \textbf{Result Variable:} The result variable is a part of the ``Ending Statement of if block". For most of the cases, the first child node of the ``Ending Statement" is the result variable.

\end{itemize}

\subsubsection{AST-walker-based Feature Extractor}
My AST-walker-based Feature Extractor tool parses the ANTLR generated parse tree (for the input Quorum code), determines the loop feature vectors, and identifies the action of a Quorum loop-if. Each of the feature vectors for a Quorum loop-if is determined from the AST as described below.

\begin{itemize}
\item \textbf{F1: Type of ending statement :} 
 
\begin{figure}[H]
 \begin{center}
    \includegraphics[width=85mm]{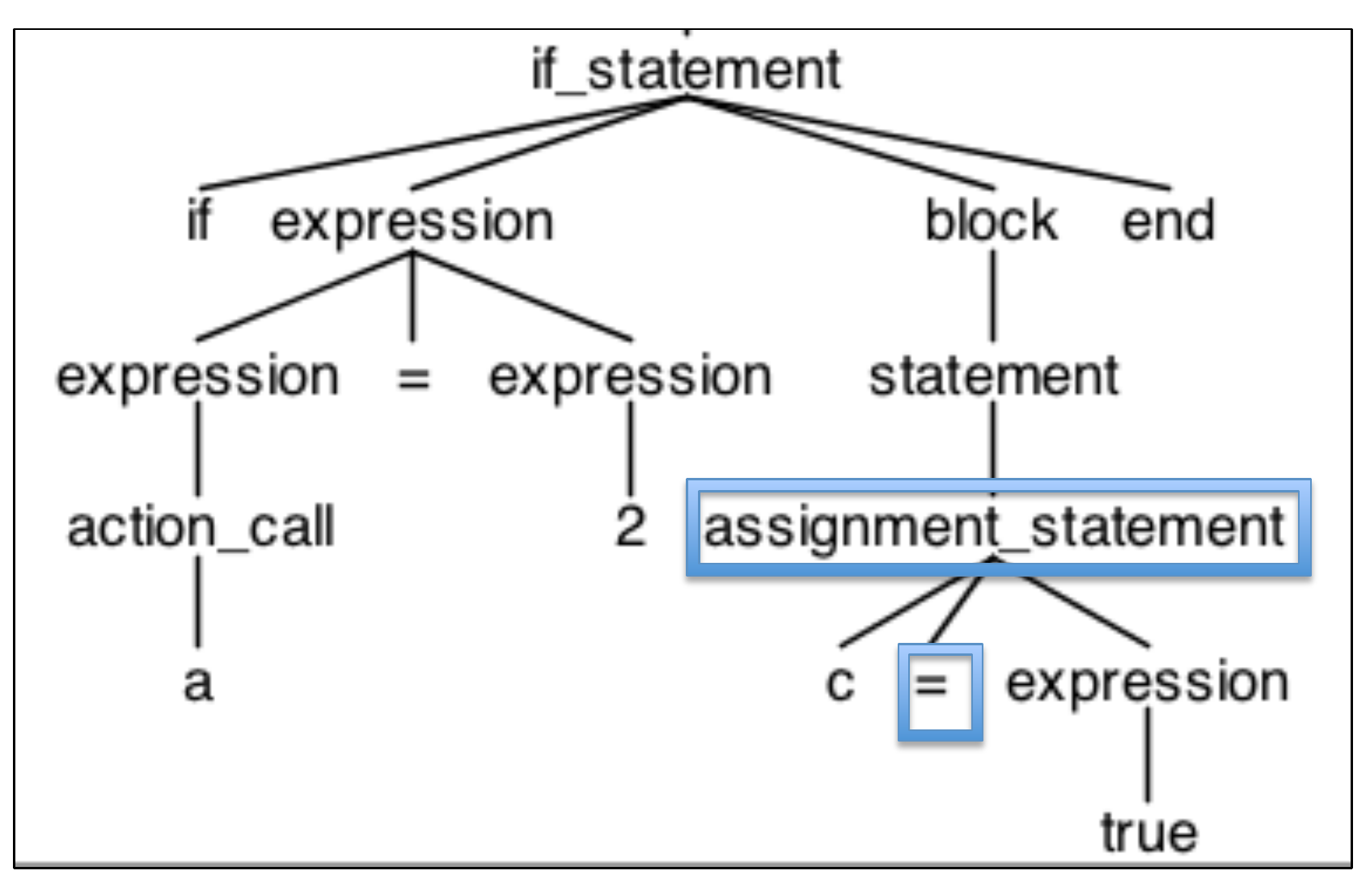}
  \end{center}
  \caption{Part of Quorum AST (F1 = 6)}
  \label{fig:F1}
\end{figure}
The Feature Extractor tool reads all the statements inside the ``if condition" of the loop-if, and identifies the ending statement of the loop, if it exists. If it does, the code reads the parse tree to find out if the statement is of type ``assignment"- assignment statements could be processed further to check if it has ``increment'' or ``decrement'' operators, or else it could be also a boolean assignment. If the ending statement is not of type ``assignment", then the code finds out whether it is of type ``solo\_method\_call" to identify a object method invocation or  a simple method call. In figure \ref{fig:F1} the block of AST shows an example of F1=6, which means the type of ending statement is boolean assignment.

\item \textbf{F2: Method name of ending statement method call :} 
\begin{figure}[H]
 \begin{center}
    \includegraphics[width=115mm]{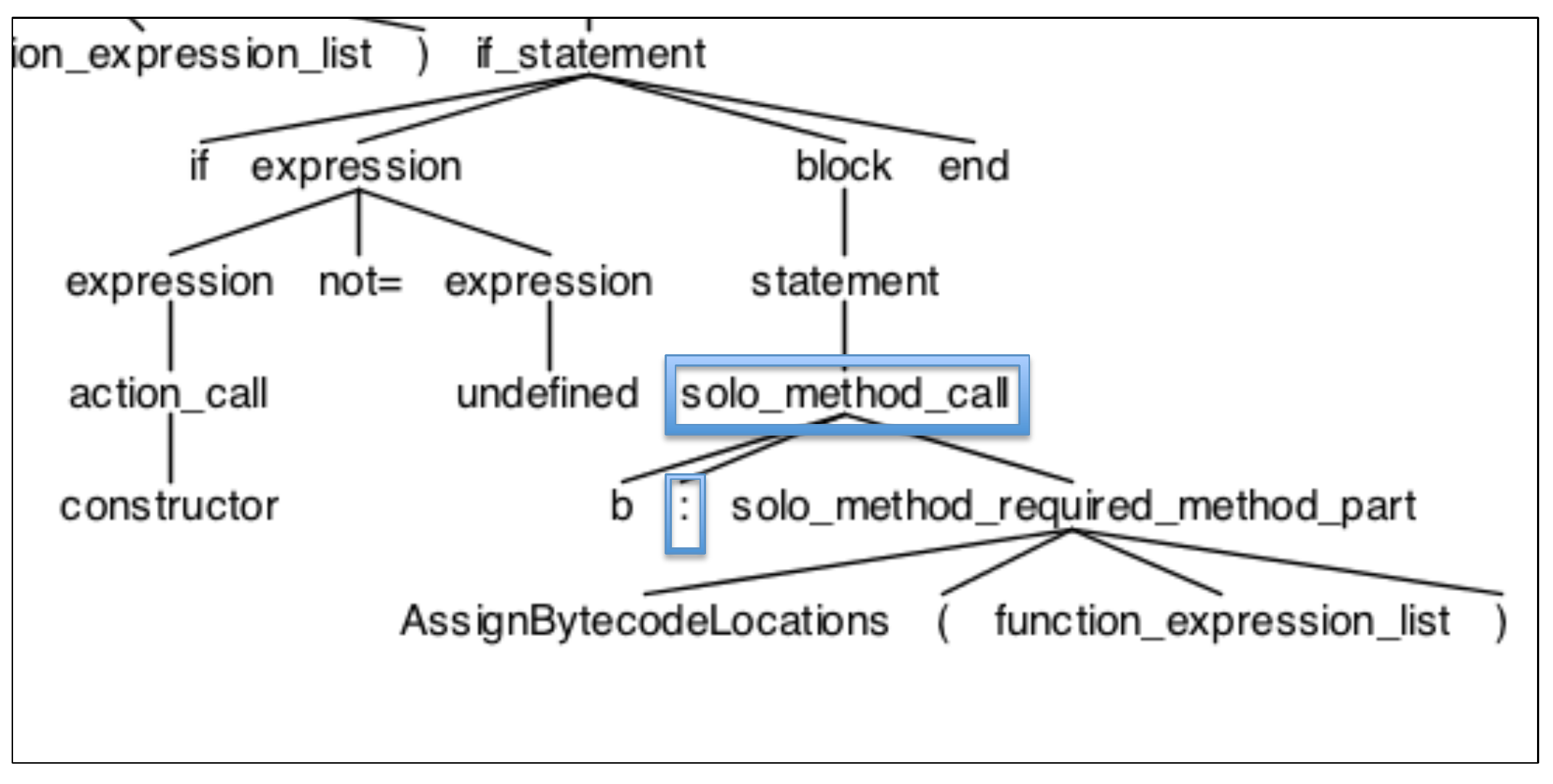}
  \end{center}
  \caption{Part of Quorum AST (F2 = " ``)}
  \label{fig:F2}
\end{figure}
If the ending statement is of type ``solo\_method\_call" , the statement is split on the operator ``:" to determine the name of the method. In figure \ref{fig:F2} the block of AST shows an example of F2=" ``, which means the method name ``AssignByteCodeLocations" of the ending statement is not on the list of methods in table \ref{table:Semantic_feature_Quorum}.

\item \textbf{F3: Elements in collection get updated :} The result variable is extracted from the child nodes of the ending statement in the parse tree. The ``expression" child node of the loop statement contains the ``loop variable". Both the result variable and the loop control variable are thus extracted, and matched to each other to find out if they are the same. The value of F3 is set to 1 if both are same, and 0 otherwise.
Please refer to figure \ref{fig:Quorum_parse_tree_gui} to locate result variable and loop control variable in an AST.

\item \textbf{F4: Usage of loop control variable in ending statement :} If the loop control variable is directly used in the ending statement or not, is already determined while extracting Feature F3. However F4, provides another option, where the loop variable might be used indirectly in the ending statement through data flow inside the loop. In this scenario, all the statements (of type: assignment or solo\_method\_call) inside the loop are scanned for possible occurrences of the loop variable or it's derived variables. If any of the derived variable occurs in the ending statement, F4 is set to 2. Please refer to figure \ref{fig:Quorum_parse_tree_gui} to locate loop control variable and the ending statement in an AST.

\item \textbf{F5: Type of loop exit statement :} 
\begin{figure}[H]
 \begin{center}
    \includegraphics[width=115mm]{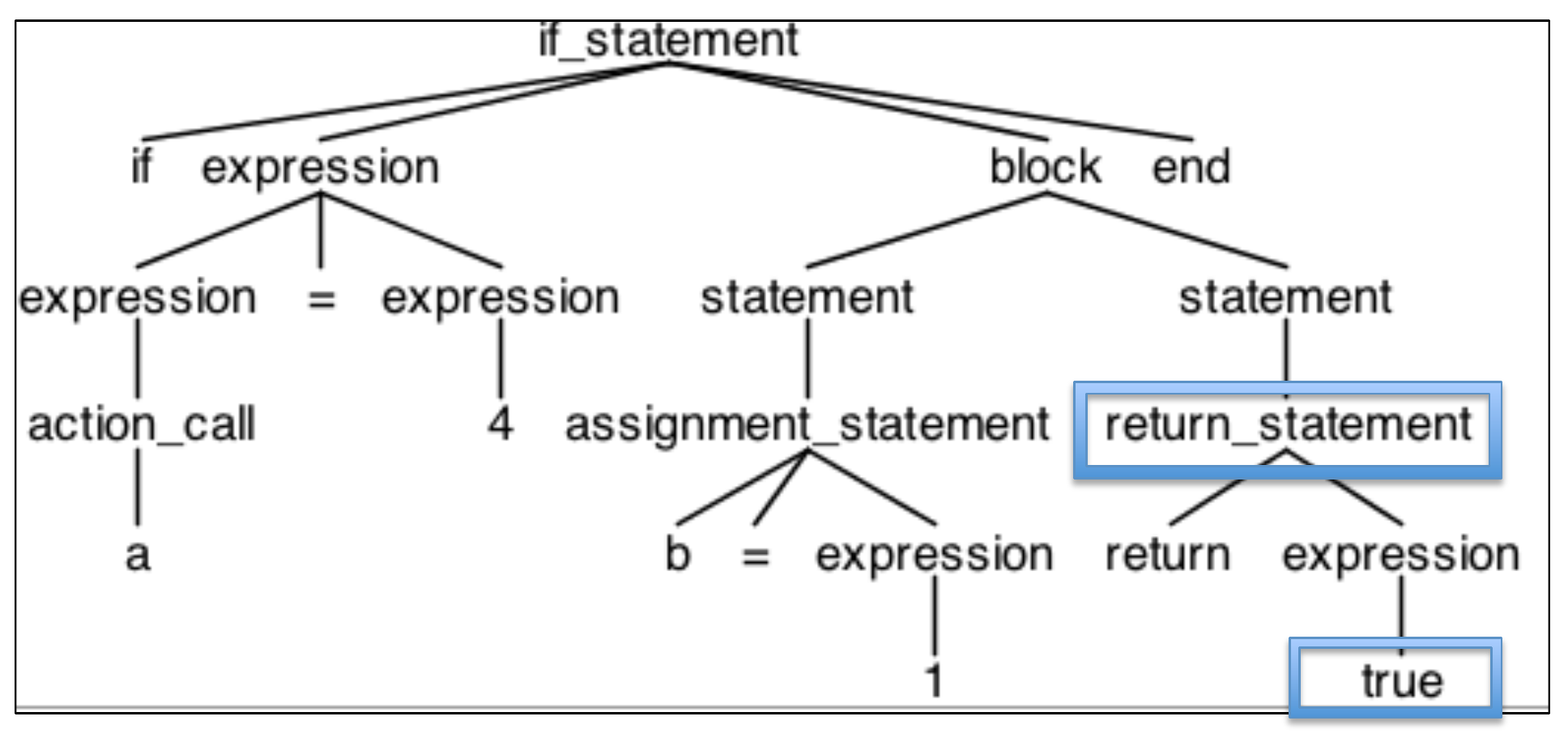}
  \end{center}
  \caption{Part of Quorum AST (F5 = 3)}
  \label{fig:F5}
\end{figure}
If the last statement inside the ``if block" is of type ``return statement" , the child nodes are further inspected to find out if the statement returns an object or a boolean value. In figure \ref{fig:F5} the block of AST shows an example of F5=3, which means the type of loop exit statement is ``return boolean".


\item \textbf{F6: Multiple collections in if condition :} 
\begin{figure}[H]
 \begin{center}
    \includegraphics[width=135mm]{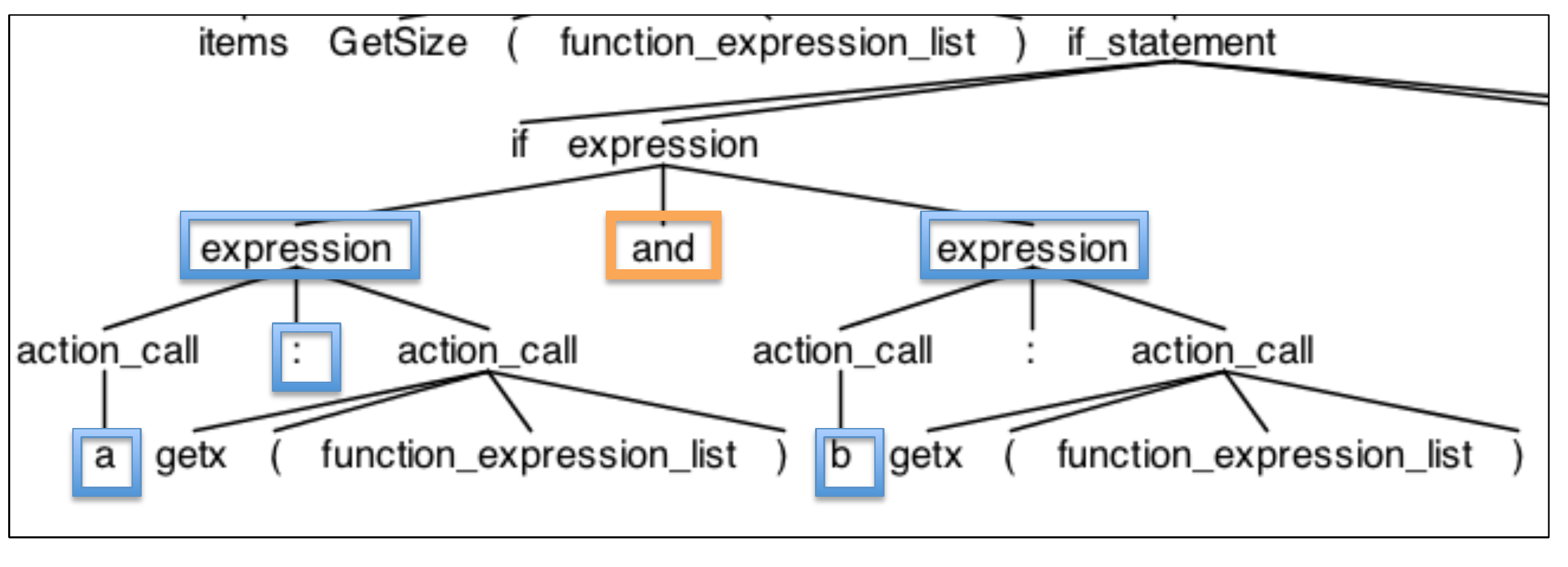}
  \end{center}
  \caption{Part of Quorum AST (F6 = 1 and F8 = 2)}
  \label{fig:F6}
\end{figure}

The tool extracts the child node ``expression" from the ``if\_statement" node, and splits it around the ``:" operator to find the object names used in the statement. F6 is set to 1, if more than one containers are used in the ``if\_statement". In figure \ref{fig:F6} the block of AST shows an example of F6=1, which means there are multiple collections (a and b) in the ``if condition".

\item \textbf{F7: Result variable used in if condition :} The ``result variable" is already been extracted to determine feature F3. The child node ``expression" of the``if\_statement" is again parsed to determine F7. Please refer to figure \ref{fig:Quorum_parse_tree_gui} to locate result variable and the ``if-condition" in an AST.

\item \textbf{F8: Type of if condition :} Identifying the operators used in the ``if\_statement" involves parsing the child node of ``expression" (which is a child node to the ``if\_statement"). The intuition for F8 is for finding max/min element in an array/collection. Having multiple clauses in a if condition is not what is required in this scenario. Hence, if more than one conditions are used in the ``if condition", F8 is simply set to 2. In figure \ref{fig:F6} the block of AST shows an example of F8=2, which means there are multiple conditions (shown by ``and") in the ``if condition".

\end{itemize} 

\vspace{0.2cm}

After the feature vectors are extracted, they are mapped to the action identification model in \cite{XL15}, and the high level actions for each of the loop-ifs are determined.

\section{Analysis of the Generality of The Loop Action Model}
\subsection{Applying the Loop Action Model for Quorum}

Based on the successful implementation of this project, it can be concluded that applying the Loop Action Model \cite{XL15} (originally designed for Java) to any other programming language, especially an object-oriented language, is feasible. Identifying a loop-if structure in another procedural programming language (like C, C++, C\#, Perl, Python, FORTRAN, MATLAB etc) is achievable, as every procedural programming language has conditional statements (if-statements) and iterative statements (loops). However it would be difficult in pure functional languages (like Haskell, etc) or machine/assembly level languages. 
\vspace{0.2cm}

Identifying the relevant pieces like loop control variable, ending statement of ``if block", etc are extremely important for determining the feature vector of a ``loop-if". However the complexity of identifying those largely depends on the structure and type of the loop and conditional constructs of the language. A for-loop statement is available in most imperative programming languages. Generally, for-loops fall into one of the following categories: Traditional for-loops (e.g. Java, C++), Iterator-based for-loops (e.g. Python), Vectorised for-loops (e.g. FORTRAN 95) and Compound for-loops (e.g. ALGOL 68) \cite{WikiFor}. While identifying the relevant pieces for vectorised for-loops would be difficult, it might be straight-forward for the rest. The conditional constructs is mostly common across many programming languages. Although the syntax varies quite a bit from language to language, the basic structure remains the same \cite{WikiIf}. 
\vspace{0.2cm}

The possible feature values of a ``loop-if" structure were designed by Wang et al. keeping in mind the characteristics of Java loop-ifs only \cite{XL15}. So some of the feature values are inapplicable to loop-ifs of other languages. For example, the feature F5 has a possible value of 5, which means the type of the loop exit statement is ``throw". But, Quorum does not support the keyword ``throw", and hence F5 = 5 is not a possible value in this case. Similarly, feature F1 has a possible value of 5, which means the type of the ending statement is ``object method invocation". Evidently, this feature value would only be possible for object-oriented languages. Also, the set of method names used as the possible values of feature F2 might not exist for other languages. However, in general, it can be concluded that feature extraction would be easier for object-oriented languages, compared to the rest. 



\subsection{Evaluation of Automatically Identified Actions in Quorum}

\subsubsection{Methodology}
To determine the potential impact of automatically identifying the high level actions of loop-ifs, we ran the action identifier on all Quorum language source code repository that was used as our test data set. Cumulatively, there are 
468 programs for the compiler and 
more than thousand for library files.  Data was gathered on the frequency of each high level action that was automatically identified. 

\subsubsection{Results}
\begin{figure}[H]
 \begin{center}
    \includegraphics[width=110mm]{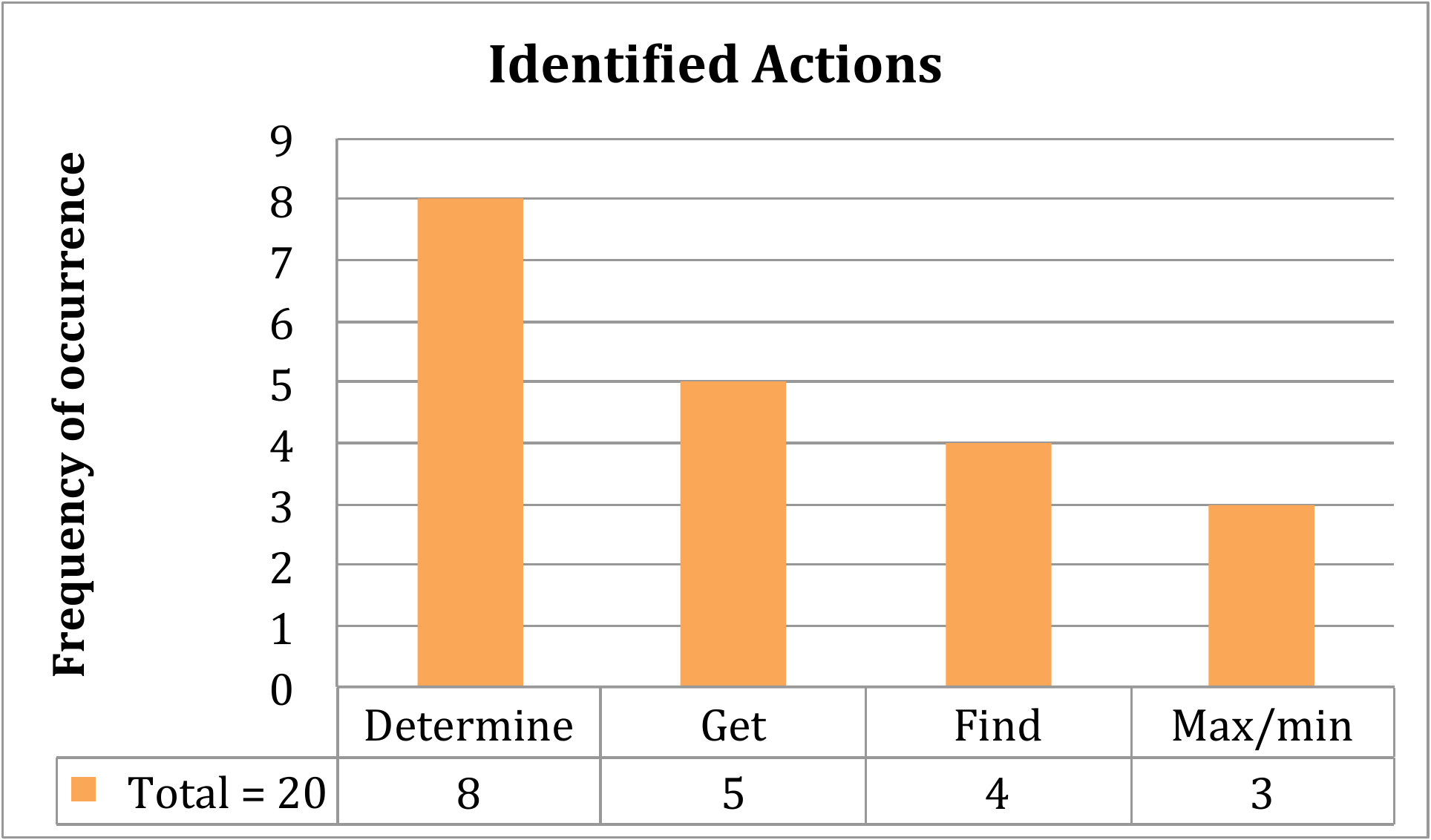}
  \end{center}
  \caption{Identified high level action distribution in Quorum Source Repository}
  \label{fig:Result_summary}
\end{figure}

In the data set, 40 loop-ifs were identified in total, out of which, high level actions were identified for 20 loop-ifs (i.e. 50\%), after mapping the feature vectors with the existent action identification model. 4 types of high level actions were identified, in the frequency of: `max/min'-(3) , `find'-(4)  , `get'-(5) , `determine'-(8) as depicted in Figure \ref{fig:Result_summary}. 

As the source code repository (compiler and standard library files) of the Quorum programming language was the only data set available for validation, measuring the correctness of the results of the automatic action identifier was straight forward. The feature vectors and the actions identified by the automatic action identifier for the 20 loop-ifs (for which the high level actions are identified) were matched against the results of the Manual Action Identification (discussed in section 3.2) of the same loop-ifs. It was found that no incorrect action is identified for any of the 20 loop-ifs , which gives a 100\% accuracy of the action identifier tool for Quorum.

\vspace{0.2cm}
For the 20 loop-ifs which could not be identified, 11 were not identified as no corresponding entries were found in the action identification model, and for the rest, actions could not be identified as the last lexical statement of the `if block' contained method calls with method names not listed in the action identification model used. This is expected as the action identification model \cite{XL15} was based on top 100 most frequent feature vectors in Java, the table does not have entries for all possible combination of feature value pairs.

\vspace{0.2cm}

Out of the 11 loop-ifs which could not be identified (for no corresponding entries) - a few were very close to identifying action `find'. 
To appreciate the problem of exploring whether the loop action model can be modified for Quorum, consider the following example Quorum loop code segment:

%

\small\begin{verbatim}
action GetStaticKey returns text
      key = ""
      i = 0
      repeat GetSize() times
          key = key + names:Get(i)
          if i < GetSize() - 1
              key = key + "."
          end
          i = i + 1
      end
      return key
  end
\end{verbatim}
\normalsize

The feature vector for the example code fragment is: ( F1:1, F2:0, F3:0, F4:2, \textbf{F5:0}, F6:0, F7:0, F8:1 ).
F1 indicates that the ending statement is an assignment. F2 indicates there is no method name for an ending statement method call. F3 indicates that no element in a collection is updated. F4 indicates that the loop control variable is indirectly used through the data flow in the ending statement. F5 indicates that there is no loop exit statement. F6 indicates that there is no collection in if condition. F7 indicates that the result variable is not used in the if condition. F8 indicates that the type of the if condition is not numeric comparison. Mapping the feature vector for this sample Quorum code with the action identification model in table \ref{table:action_id_model}, loop action for the particular loop cannot be identified because no corresponding entry was found in the action identification model. However, the feature vectors in table \ref{table:action_id_model} for identifying action find are ( F1:1, F2: , F3: , F4:1,2, \textbf{F5:1,2,4}, F6: , F7: , F8: ) and ( F1:0, F2: , F3: , F4: , F5:4, F6: , F7: , F8: ). It is noted that the feature vector for this sample Quorum code is very close to the first feature vector for the action find, except that F5 is 0, not 1,2 or 4. 
So as future work, there is a potential of modifying the action identification table according to the specifics of Quorum language, for better effectiveness.

\subsubsection{Comparing the results: Java vs. Quorum}
Keeping in mind the small size of data set currently available for research to implement the model for Quorum, the results look quite promising when compared to Java. The percentage of automatically identified high level actions in both the languages is close to 50\%, and accuracy of the model implemented in Quorum is 95\%, which seems to be better than what we had for Java which was determined through non-author human judgements. The number of types of identified loop actions in Quorum is low, which might be accounted to the fact that the model was designed keeping in mind the most frequent loop actions occurring in Java. The top 3 frequently identified loop actions are still the same for both the programming languages, only in a different order.

\begin{table}[H]
\scriptsize
\begin{center}
\begin{tabular}{ |p{5cm}|p{5cm}|p{5cm}| } 
 \hline
 \textbf {Category} & \textbf{Java} & \textbf{Quorum} \\ 
 \hline
Data set & 7,159 open source projects &  468 compiler source code programs \\ 
 \hline
Number of identified loop-ifs & 337,294 & 40  \\ 
 \hline
Number of automatically identified high level actions & 195,277 (i.e., 57.9\%) & 20 (i.e., 50\%) \\
 \hline
Types of identified high level actions & 12 & 4\\
 \hline
Accuracy & 93.9\% & 100\% \\
 \hline
Most Frequent identified loop actions (in descending order) & Find, Determine, Get & Determine, Get, Find \\
 \hline
\end{tabular}
\caption{Comparing Results: Java (left) and Quorum(right)}
\label{table:Results}
\end{center}
\end{table}

\subsection{Threats to Validity}
The results obtained for the Quorum language source code might differ when tested on larger data sets. Also, given the code is written by a handful of developers, there is less variety of loop implementations to test, as different programmers have different coding styles. To mitigate this, as many as available Quorum loops were collected for this project. These loops are real code examples used in writing the Quorum compiler and libraries, and are not some random sample code snippets.

\section{Related Work}
The first general and extensible approach to automatically identifying action units and abstracting them as high level action phrases without manually creating templates was developed by Wang et al. \cite{XL15}. The closest work is by Sridhara et al. who automatically generated high-level actions within methods \cite{sridhara2011automatically}, by using a small set of templates that were developed by manually examining code.

\vspace{0.2cm}
	
	This project is related to generating internal comments for the identified high level actions. Most comment generation work is focused on creating summaries for methods or classes \cite{sridhara2010towards} \cite{moreno13}. However, Wong et al. mine question and answer sites for automatic comment generation \cite{6693113}. They extract code-description mappings from the question title and text, use heuristics to refine the descriptions, and use code clone detection to find source code snippets that are almost identical to the code-description mapping. 

\section{Conclusion and Future Work}
Application of the Loop Action Model for Quorum demonstrated the feasibility of implementing the same model to other programming languages apart from Java. Building a tool for automatic identification of high level loop actions shows the potential to generate internal comments for loop structures and thus help programmers to save time and effort in apprehending code. Also, beyond an approach to generating internal comments, tools that rely on the words in the source code and comments for analysis will benefit when the high level action words do not already appear in the loop code (e.g., search tools, comment generator tools).

\vspace{0.2cm}

Based on the success of this approach, it is worth exploring the applicability of the approach to other programming languages apart from Java and Quorum. It will also be interesting to investigate how the automatic tool works on unseen Quorum code written by different  programmers, and determine the accuracy and types of high level actions occurring on a larger data set.

\newpage

\bibliographystyle{plain}


\end{document}